\documentclass[preprint,12pt]{elsarticle}

\usepackage{amsmath}
\usepackage{amsfonts}
\usepackage{amssymb}
\usepackage{fancyhdr}
\usepackage{calc}
\usepackage{url}
\usepackage{tabularx}
\usepackage{graphicx}
\usepackage[usenames,dvipsnames,table]{xcolor}
\usepackage{makeidx}
\usepackage{wrapfig}
\usepackage{subcaption}
\usepackage{hyperref}
\usepackage{url}
\usepackage{verbatim}
\usepackage{algorithm}
\usepackage{algorithmic}

\usepackage[normalem]{ulem}

\begin{document}

\begin{frontmatter}

\title{Performance comparison of CFD-DEM solver MFiX-Exa, on GPUs and CPUs}

\author[cuboulder1]{Shandong Lao}
\ead{shandong.lao@colorado.edu}
\author[cuboulder1]{Aaron Holt}
\author[nreladdesif]{Deepthi Vaidhynathan}
\author[nreladdesif]{Hariswaran Sitaraman}
\author[cuboulder2]{Christine M. Hrenya}
\author[cuboulder1]{Thomas Hauser}
\address[cuboulder1]{Research Computing, University of Colorado, Boulder}
\address[nreladdesif]{Computational Science Center, National Renewable Energy Laboratory}
\address[cuboulder2]{Chemical and Biological Engineering, University of Colorado, Boulder}

\begin{abstract}
We present computational performance comparisons of gas-solid simulations performed on 
current CPU and GPU architectures using MFiX Exa, a CFD-DEM solver that leverages hybrid 
CPU+GPU parallelism. A representative fluidized bed simulation with varying particle numbers 
from 2 to 67 million is used to compare serial and parallel performance. A single GPU was observed to be about 10 times faster compared to a single CPU core. The use of 3 GPUs on a single compute node was observed to be 4x faster than using all 64 CPU cores. We also observed that using an error controlled adaptive time stepping scheme for particle advance provided a consistent 4x speed-up on both CPUs and GPUs. Weak scaling results indicate superior parallel efficiencies when using GPUs compared to CPUs for the problem sizes studied in this work.
\end{abstract}

\end{frontmatter}


\section{Introduction}
Gas-solid multiphase systems are ubiquitous in several applications such as pharmaceutical, mining and petrochemical industry. The Computational fluid dynamics (CFD)-Discrete element method (DEM) is a well-established theoretical approach for modeling granular phenomenon in these systems. The motion of individual solid particles are resolved in the CFD-DEM approach along with a continuum description for the background gaseous phase. There are minimal modeling assumptions compared to two-fluid \cite{passalacqua2011implementation,rahimi2018computational} or mixture models \cite{sitaraman2015multiphysics, sitaraman2019coupled} in this approach which are mainly associated with particle drag and inter-particle collisions. CFD-DEM is therefore a rigorous method which is amenable to small-scale systems with $\sim$ few million particles while its application to large-scale industrial systems ($\sim$ few billion particles) \cite{liu2014challenges,gan2016,cocco2017} is constrained by high computational requirements.
Recent advances in high performance computing (HPC) architectures with multiple Central Processing Units (CPU) cores and Graphics Processing Units (GPU) acceleration provide a viable pathway to perform large-scale CFD-DEM simulations. Several CFD-DEM solvers have been ported to parallel computing frameworks \cite{gopalakrishnan2013development,kafui2011parallelization,pozzetti2018co,dufresne2020massively} using the message passing interface (MPI) and exhibit improved performance. CFD-DEM coupling although can impose conflicting configurations \cite{sitaraman2016balancing,pozzetti2018co} for parallel domain decomposition for which dual grids or unified load-balancing algorithms may be necessary. DEM has received considerable attention with regard to use of GPU acceleration due to the simplified nature of particle advance \cite{qi2015gpu,gan2016,tian2017implementing}, while CFD using traditional finite-volume methods most often require linear-solvers that impose algorithmic constraints on extracting GPU performance. Therefore, CFD solvers that are efficient on CPUs have been coupled with accelerator-enhanced DEM solvers \cite{kloss2012models,he2020cpu}. Linear solvers that use scalable multi-grid \cite{basu2017compiler} and Krylov Subspace methods \cite{matsumoto2019implementation} have now been demonstrated to show superior performance on GPU accelerators. MFiX-Exa, the software employed in this work, is an emerging CFD-DEM solver that utilizes GPU acceleration for both continuum and particle solves within a hybrid MPI+GPU parallelism framework. The performance implications for such a novel framework requires careful analysis of computational performance with respect to scalability, memory utilization and arithmetic intensity over hybrid HPC architectures, which is the main contribution of this work.

This paper is organized as follows. We first provide a brief description of the numerical methods and the parallelization algorithms within MFiX-Exa. We then present comparison of compute and memory performance on CPU and GPU architectures for a physically relevant fluidized bed simulation. Finally we show parallel performance in weak-scaled settings that utilize the hybrid MPI+GPU paradigm.

\section{Numerical methods and algorithms}
CFD-DEM solver, MFiX-Exa \cite{musser2021mfix,syamlal1993}, that leverages software library, AMReX \cite{zhang2019}, is used in this work. AMReX is a performance portable library specifically tailored towards massively parallel, block-structured adaptive-mesh-refinment (AMR) based applications and has been actively employed in several recent scalable frameworks in combustion \cite{sitaraman2021adaptive}, wind \cite{amrwind} and astrophysics \cite{zingale2018meeting}. The equations for gas-phase mass and momentum are solved in the incompressible form for the fluid update while the particle position, translational and angular velocities are updated  using Newton's laws of motion. Readers are directed to references \cite{fullmer2019,mfixexa_doc,sitaraman2020error,musser2021mfix} for details on formulation and numerical methods. 
A brief description of the fluid-particle update algorithm in a typical CFD-DEM time-step where particle collision time-scales are smaller than fluid time-scales is described below: 
 \begin{enumerate}
    \item{The mass and momentum equations for the fluid phase are advanced using an incompressible projection scheme for a time-step of $\Delta t_f$}
\item{Particles are advanced using multiple sub-iterations with time-step $\Delta t_p$ for time $\Delta t_f$.}
\begin{enumerate}
    \item{Particle drag is computed using interpolated fluid variables.}
    \item{Collisional forces are computed using particle neighbor-lists.}
    \item{Particle velocity and position are advanced using an explicit scheme.}
\end{enumerate}
\item{Solids volume fraction and momentum transfer terms are computed from particle data on the fluid grid using volume filtering.}
\end{enumerate}
\begin{figure}
\centering
  \includegraphics[width=.9\textwidth]{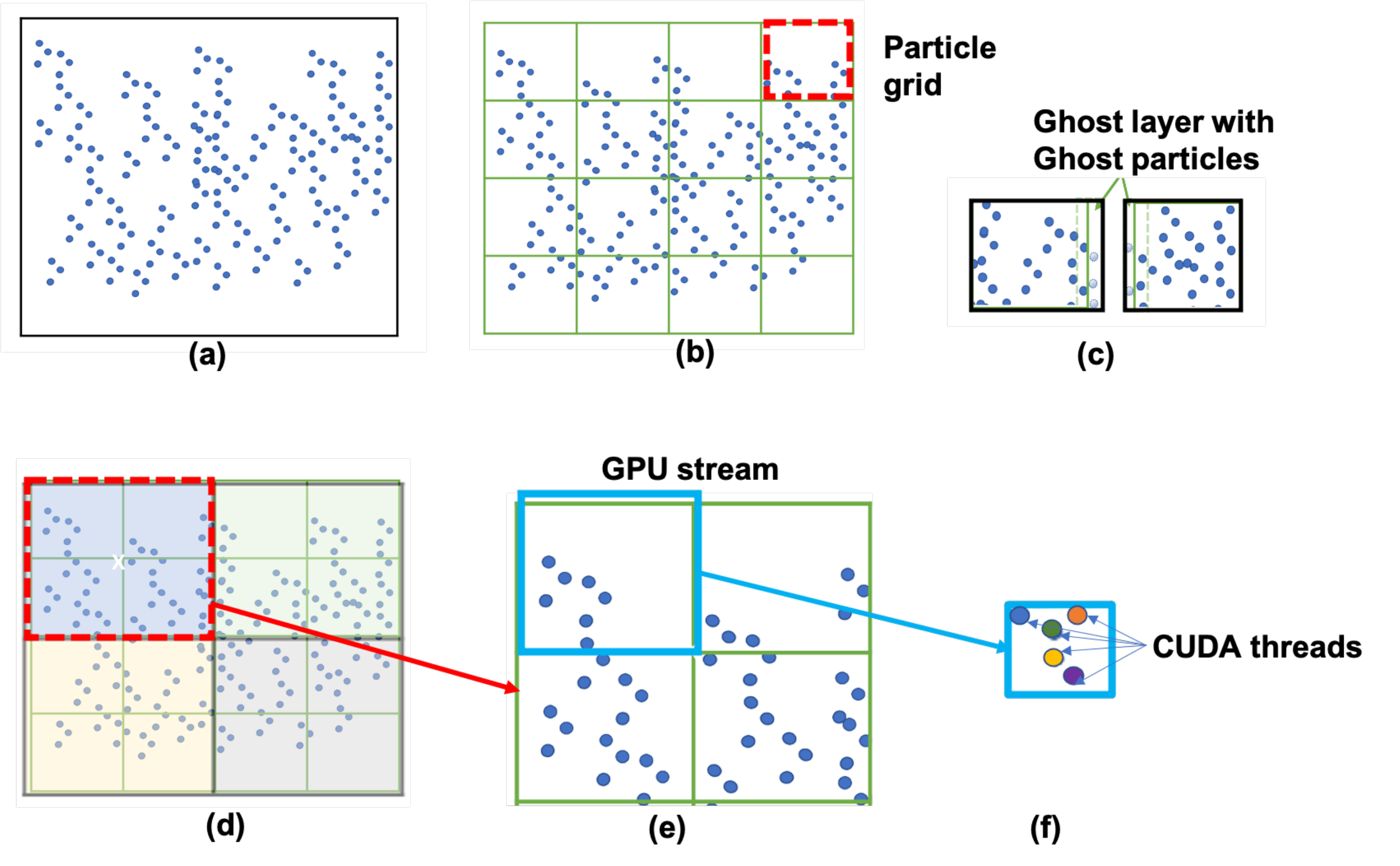}
\caption{(a) shows an example distribution of particles in a Cartesian domain, (b) shows domain decomposition into 16 \emph{boxes} with 4 boxes along each direction,  (c) shows two adjacent \emph{boxes} with \emph{ghost}-layers that are used for communication of particles between MPI ranks, (d) shows an example distribution of 16 \emph{boxes} among 4 MPI ranks, each rank owns 4 \emph{boxes}, (e) shows the assignment of the 4 \emph{boxes} local to an MPI rank to multiple GPU streams and (f) shows the allocation of particles in a stream to GPU threads.}
\label{domdecomp-fig}
\end{figure}
A hybrid distributed (MPI) and shared (GPU) memory  paradigm is used in the solver for parallelization. The domain decomposition approach is as shown in Fig.\  \ref{domdecomp-fig}.  The computational domain that uses an uniform Cartesian grid is divided into sub-domains or \emph{boxes} with co-located particles, as shown in Fig.\ \ref{domdecomp-fig}(b). A space-filling curve \cite{farooqi2018phase} algorithm is used to distribute the \emph{boxes} among processors. The number of \emph{boxes} typically is equal or greater than the number of MPI ranks, as indicated by an example decomposition in Fig.\ \ref{domdecomp-fig}(d). Shared memory parallelism is achieved by using GPU threads that operate on one or more particles (for DEM) or fluid-cells (for CFD) within each \emph{box}, as shown in Fig.\ \ref{domdecomp-fig}(e) and (f).  An important aspect of this approach is the need to store \emph{ghost} particles at the boundary of each \emph{box}. This is shown in Fig.\ \ref{domdecomp-fig}(c) for an example case of two adjacent \emph{boxes}. The \emph{ghost} particles are stored within a single layer of cells and are updated every particle time-step among all \emph{boxes} and processors. Particles that move outside of a box are redistributed to MPI ranks that own adjacent \emph{boxes}. Particle collision forces are computed from interactions within a sphere of influence that is six times the particle radius. Potential collision partners for a given particle are stored in a neighbor-list data structure that contain partner identifiers. This data-structure is dynamically updated during particle advance after every few time-steps.

\section{Results}
\subsection{Problem description}
\begin{figure}
\centering
\includegraphics[width=.95\textwidth]{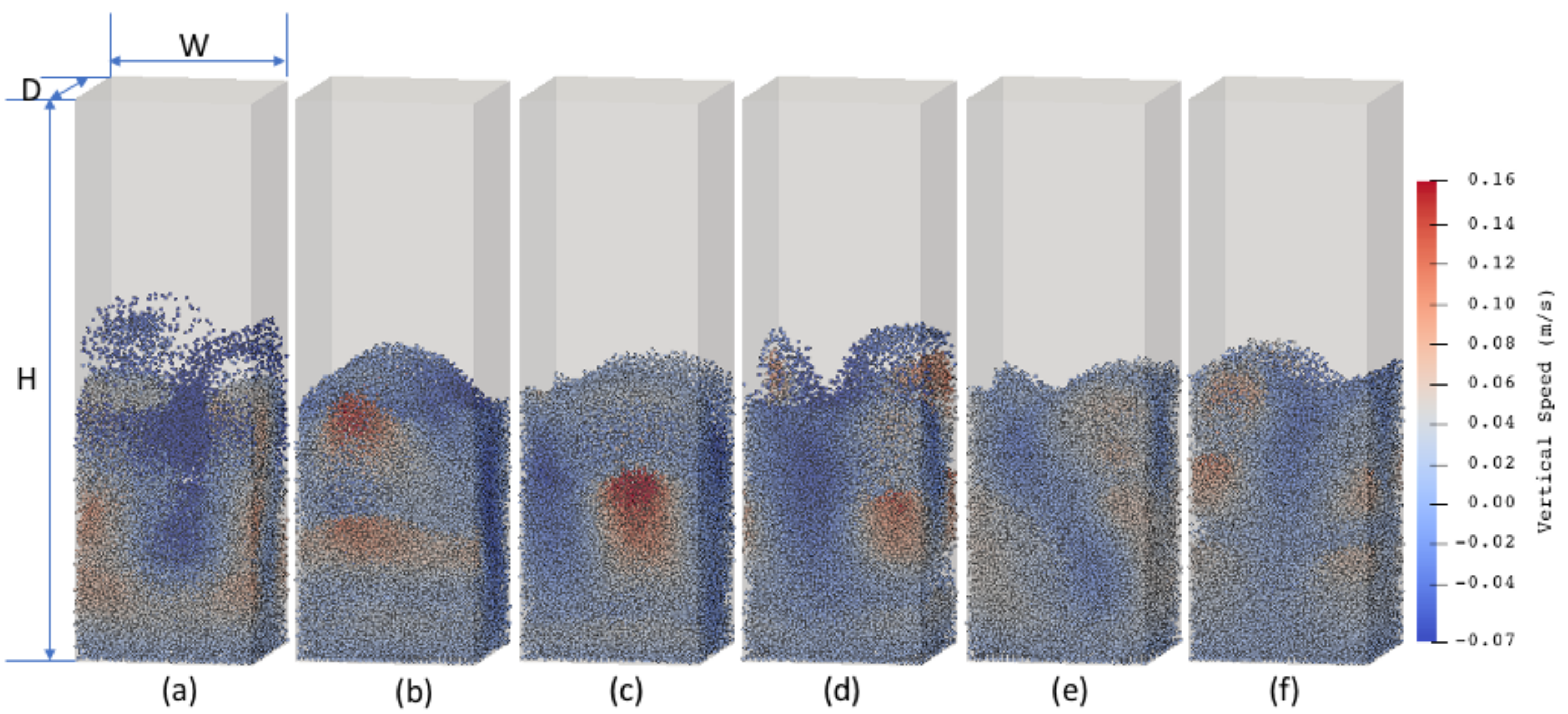}
\caption{Fluidized bed geometry (height H=0.01 m, width W=0.0032 m and depth D=0.0032 m) and particle axial speed transients at statistical steady-state shown at times (a) t=0.152, (b) t=0.381, (c) t=0.964, (d) t=1.290, (e) t=1.645, and (f) t=1.823 seconds, showing bubbling fluidization behavior.}
\label{fluidbed-fig}
\end{figure}

A rectangular fluidized bed simulation setup is used to analyze the performance of MFiX-Exa on both CPU and GPU in this study. The geometry and particle transients at statistically stationary state  are illustrated in Fig.\ \ref{fluidbed-fig} for a representative run with 40,000 particles. A series of simulations of different domain sizes and particle numbers are carried out to analyze computational performance on CPUs and GPUs. The range of domain sizes and particle numbers are listed in Table \ref{tab:geometry}. Tables \ref{tab:fluid} and \ref{tab:particle} define the fluid and solid properties used in our simulations,  respectively. 
\begin{table}
\centering
\begin{tabular}{l|r} 
 \hline
 \hline
 Parameter                    & Range \\
 \hline
 \hline
 H ($m$)                      & 0.0100 - 0.6400 \\ 
 W ($m$)                      & 0.0032 - 0.1024 \\ 
 D ($m$)                      & 0.0032 - 0.0256  \\ 
 Solid Volume fraction ($\%$) & 20.0 - 50.0 \\
 Number of Particles          & 40,000 - 67,000,000 \\ 
 \hline
\end{tabular}
\caption{Ranges of domain Size and particle numbers for the fluidized bed cases.} 
\label{tab:geometry}
\end{table}
\begin{table}
\centering
\begin{tabular}{l|r}
\hline
\hline
 Parameter                          & Value \\
 \hline
 \hline
 Viscosity ($kg$·$m^{-1}$ $s^{-1}$) & $2.0 \times 10^{-5}$ \\ 
 Density ($kg/m^3$)                 & 1.000 \\ 
 Gas inlet Velocity ($m/s$)         & 0.015 \\
 Drag model                         & BVK\cite{beetstra2007}  \\
 \hline
\end{tabular}
\caption{Fluid properties used in the fluidized bed simulations.} 
\label{tab:fluid}
\end{table}
\begin{table}
\centering
\begin{tabular}{l|r} 
 \hline
 \hline
Parameter                                        & Value \\
 \hline
 \hline
 Diameter ($m$)                                  & $1.0 \times 10^{-4}$ \\ 
 Density ($kg/m^3$)                              & 1000.00 \\ 
 Spring Constant ($N/m$)                         &   10.00 \\
 Sliding Friction Coefficient                    &    0.00 \\
 Particle-Particle Restitution Coefficient       &    0.80 \\
 Particle-Wall Restitution Coefficient           &    1.00 \\
 Tangential-to-Normal Spring Constant Factor     &    0.28 \\
 Tangential-to-Normal Damping Coefficient Factor &    0.50 \\
 \hline
\end{tabular}
\caption{Particle properties used in the fluidized bed simulations.} 
\label{tab:particle}
\end{table}
\subsection{Compute architectures}
Serial and single node benchmarking and profiling was completed on University of Colorado Research Computing’s \emph{Blanca} CPU+GPU nodes. Each \emph{Blanca} CPU+GPU node contains two AMD EPYC 7452 processors and three Nvidia A100 PCIe cards each with 40 GB of GPU device memory. The hardware on these nodes are more recent and have more resources per node allowing for larger problem sizes. Additionally, hardware counters were enabled on all of these nodes allowing for more in-depth profiling such as roofline analysis.  A summary of \emph{Blanca}'s hardware is as shown below in Table \ref{tab:blancagpu}. MFiX-Exa was compiled on the \emph{Blanca} using GCC 8.2.0, OPENMPI 4.0.0, and CUDA 11.2 compilers. However, The \emph{Blanca} CPU+GPU nodes only have a 10 GB ethernet connecting them, which may reduce the performance of multi-node computational jobs.
\begin{table}
\centering
\begin{tabular}{l|r} 
 \hline
 \hline
 Hardware & Specification \\
 \hline
 \hline
 CPU & 2x Dual AMD EPYC 7452 \\ 
 Total Cores & 64 \\ 
 Total Threads & 128 \\ 
 L3 Cache Per Core & 128 MB \\ 
 Memory & 264 GB \\
 GPU & 3x NVIDIA A100 PCIe \\
 CUDA Cores & 6912 \\
 GPU Memory & 40 GB \\
 \hline
\end{tabular}
\caption{\emph{Blanca} CPU+GPU node hardware} 
\label{tab:blancagpu}
\end{table}
\begin{table}
\centering
\begin{tabular}{l|r} 
 \hline
 \hline
 Hardware & Specification \\
 \hline
 \hline
 CPU & Dual Intel Xeon Gold Skylake 6154 \\
 Total Cores & 36 \\ 
 Total Threads & 36 \\ 
 L3 Cache Per Core & 75 MB \\ 
 Memory & 96 GB \\ 
 \hline
\end{tabular}
\caption{\emph{Eagle} CPU only node hardware} 
\label{tab:eaglecpu}
\end{table}
\begin{table}
\centering
\begin{tabular}{l|r} 
 \hline
 \hline
 Hardware & Specification \\
 \hline
 \hline
 CPU & Dual Intel Xeon Gold Skylake 6154\\
 Total Cores & 36 \\ 
 Total Threads & 36 \\ 
 L3 Cache Per Core & 75 MB \\ 
 Memory & 768 GB \\
 GPU & 2x NVIDIA Tesla V100 PCIe \\
 CUDA Cores & 5120 \\
 GPU Memory & 16 GB \\
 \hline
\end{tabular}
\caption{\emph{Eagle} CPU+GPU node hardware} 
\label{tab:eaglegpu}
\end{table}
Multi-node performance benchmarking was completed on the National Renewable Energy Laboratory (NREL)'s supercomputer, \emph{Eagle}. Both a CPU only node and a CPU+GPU node were used for benchmarking runs. Both compute node types use two Intel Xeon Gold Skylake 6154 processors and are connected by a 100 GB/s EDR Infiniband network. The CPU only nodes have 96 GB of memory whereas the CPU+GPU nodes have 768 GB. The CPU+GPU nodes have two NVIDIA Tesla V100 PCIe cards with 16 GB of device memory for each one of them. Tables \ref{tab:eaglecpu} and \ref{tab:eaglegpu} give a hardware summary of the \emph{Eagle} compute nodes used in this work.  MFiX-Exa was compiled using GCC 8.4.0, OpenMPI 4.0.4 and CUDA 10.2 compilers on \emph{Eagle}. 
\subsection{Serial and Single Node performance}
%
%
\begin{figure}
\centering
\begin{subfigure}{0.85\textwidth}
\includegraphics[width=.99\textwidth]{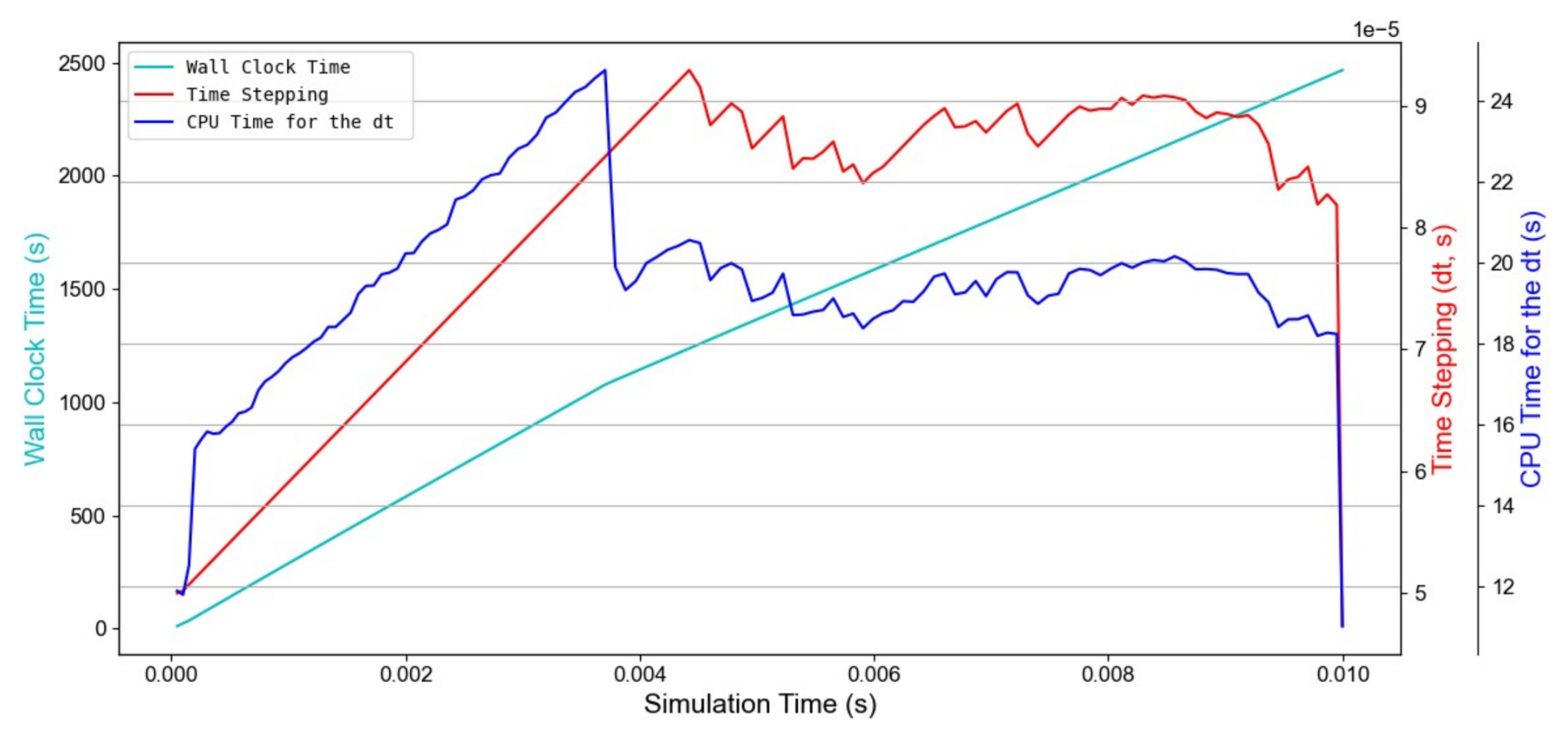}
\caption{}
\end{subfigure}
\begin{subfigure}{0.85\textwidth}
\includegraphics[width=.99\textwidth]{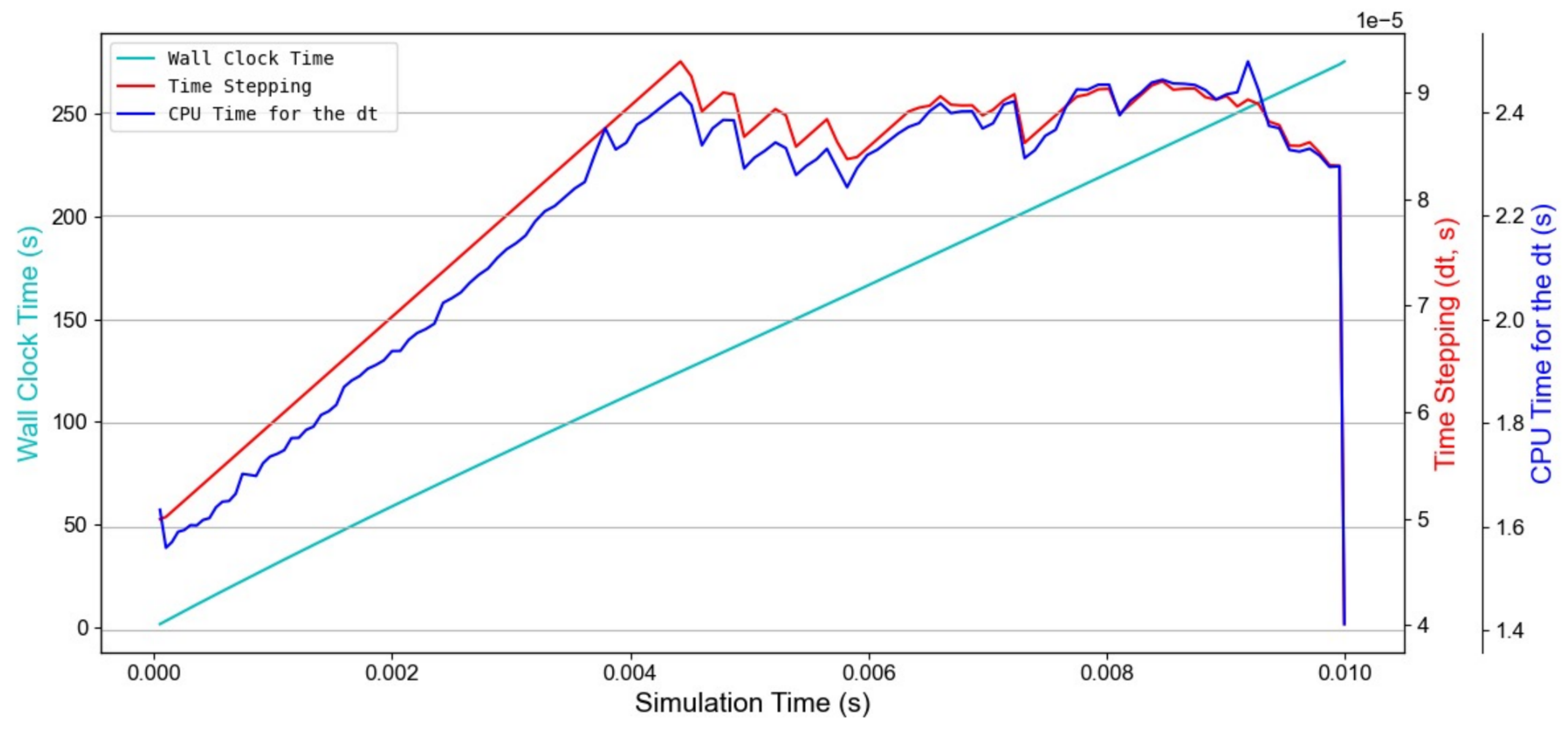}
\caption{}
\end{subfigure}
\caption{Time history of time-step size and costs per time-step for a 160,000 particle fluidized bed simulation performed on (a) single CPU core and (b) a single GPU, respectively.}
\label{flubed-160k}
\end{figure}
\subsubsection{Serial performance variation over time steps}
A single V100 graphics card on \emph{Eagle} ran a 160,000 particle simulation on a $32 \times 50 \times 32$ grid for 10 ms in 270 seconds which is $\sim$ 10x faster than a single Intel Skylake CPU that took 2400 seconds to completion as shown in Fig.\ \ref{flubed-160k}(a) and (b). These figures also demonstrate the step performance during the early transients and at statistical steady state. Both the CPU and GPU runs follow a similar pattern in terms of fluid update time-step size over the course of the simulation. The main difference being the GPU case climbed to a time per step of 2.4 seconds at simulation time of 0.004 before stabilizing whereas the CPU case hit a time per step of 24 seconds, dropped to 19 seconds and then stabilized. Overall, the single GPU case was expected to be faster than a single CPU core case because a multi-threaded V100 GPU has higher peak performance compared to a single CPU core.
\begin{table}
\centering
\begin{tabular}{l|c|r} 
 \hline
 \hline
Problem size factor & Domain $W{\times} H{\times}D\ (m^{3})$   & Particle no: ($\times 10^{6}$) \\
 \hline
 \hline
 \ 1x  &  $0.0128\times0.02\times0.0128$  &  2.10 \\ 
 \ 2x  &  $0.0128\times0.04\times0.0128$  &  4.20 \\ 
 \ 4x  &  $0.0128\times0.08\times0.0128$  &  8.40 \\
 \ 8x  &  $0.0128\times0.16\times0.0128$  & 16.76 \\
 16x   &  $0.0128\times0.32\times0.0128$  & 33.52 \\
 32x   &  $0.0128\times0.64\times0.0128$  & 67.03 \\
 \hline
\end{tabular}
\caption{Domain size scaling and associate particle numbers for performance studies. The Cartesian domain for the 1x case is a $64 \times 100 \times 64$ grid, and is decomposed into $8 \times 100 \times 8$ \emph{boxes}. The number of cells along the axial direction in the domain and \emph{boxes} are proportionally increased with increasing problem size. } 
\label{tab:problem-sizes}
\end{table}
\begin{figure}
\centering
\includegraphics[width=.95\textwidth]{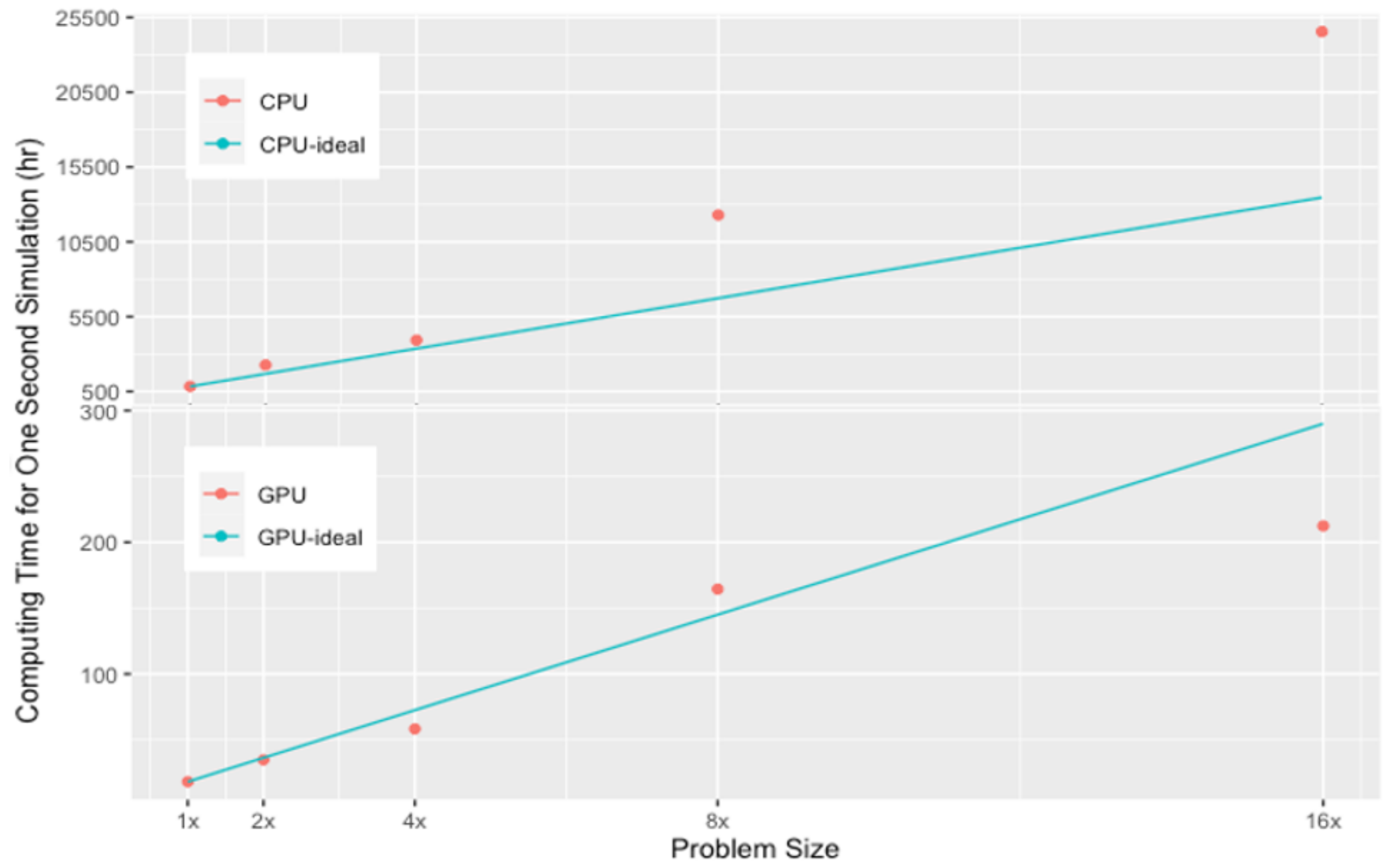}
\caption{Serial performance on a single CPU core and a GPU with varying problem sizes.}
\label{serial_sizes-fig}
\end{figure}
\subsubsection{Serial performance with problem size variation}
We compare serial performance on \emph{Blanca} with varying problem sizes in this section. For this purpose, a series of simulations with varying domain sizes and particle numbers were carried out. The domain sizes and particle numbers for these experiments are listed in Table \ref{tab:problem-sizes}. The serial performance is shown in Figure \ref{serial_sizes-fig}. The computational time increases with larger problem sizes in both CPU and GPU cases. The wall-clock times are higher than the ideal scaling for the CPU mainly due to the increased memory requirements for larger problem sizes. On the other hand, the GPU wall clock times tend to be lower than the ideal scaling (25\% lower wall clock time at 16x problem size compared to ideal) indicating efficient vectorization and memory utilization among multiple threads on the GPU for larger problem sizes. The GPU runs are about 10x faster than the serial CPU run and is consistent across varying problem sizes. 
\begin{figure}
\centering
\includegraphics[width=.95\textwidth]{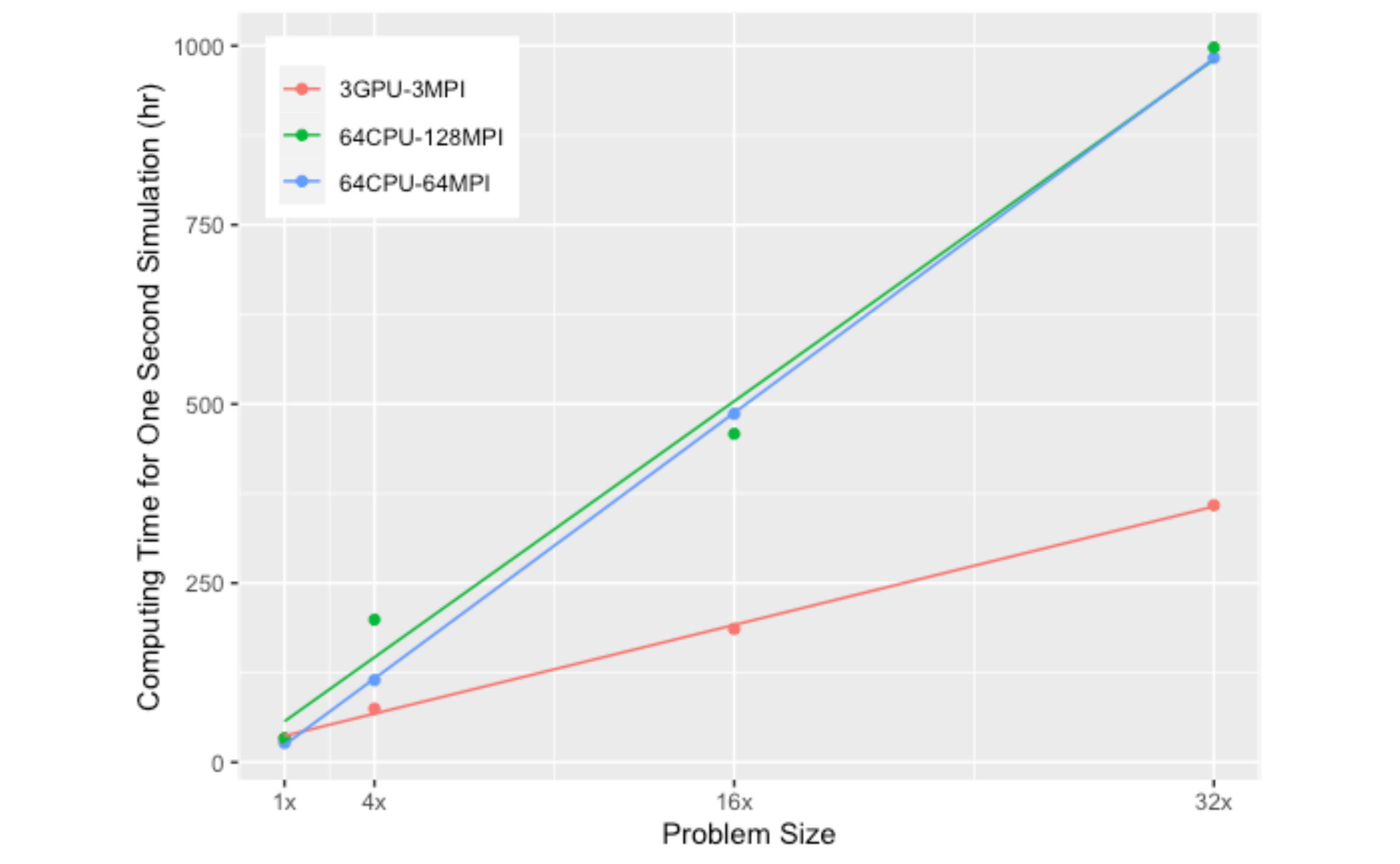}
\caption{Simulation Performance for a single node cases with varying problem sizes.}
\label{one_node-fig}
\end{figure}
\begin{figure}
\centering
\includegraphics[width=.95\textwidth]{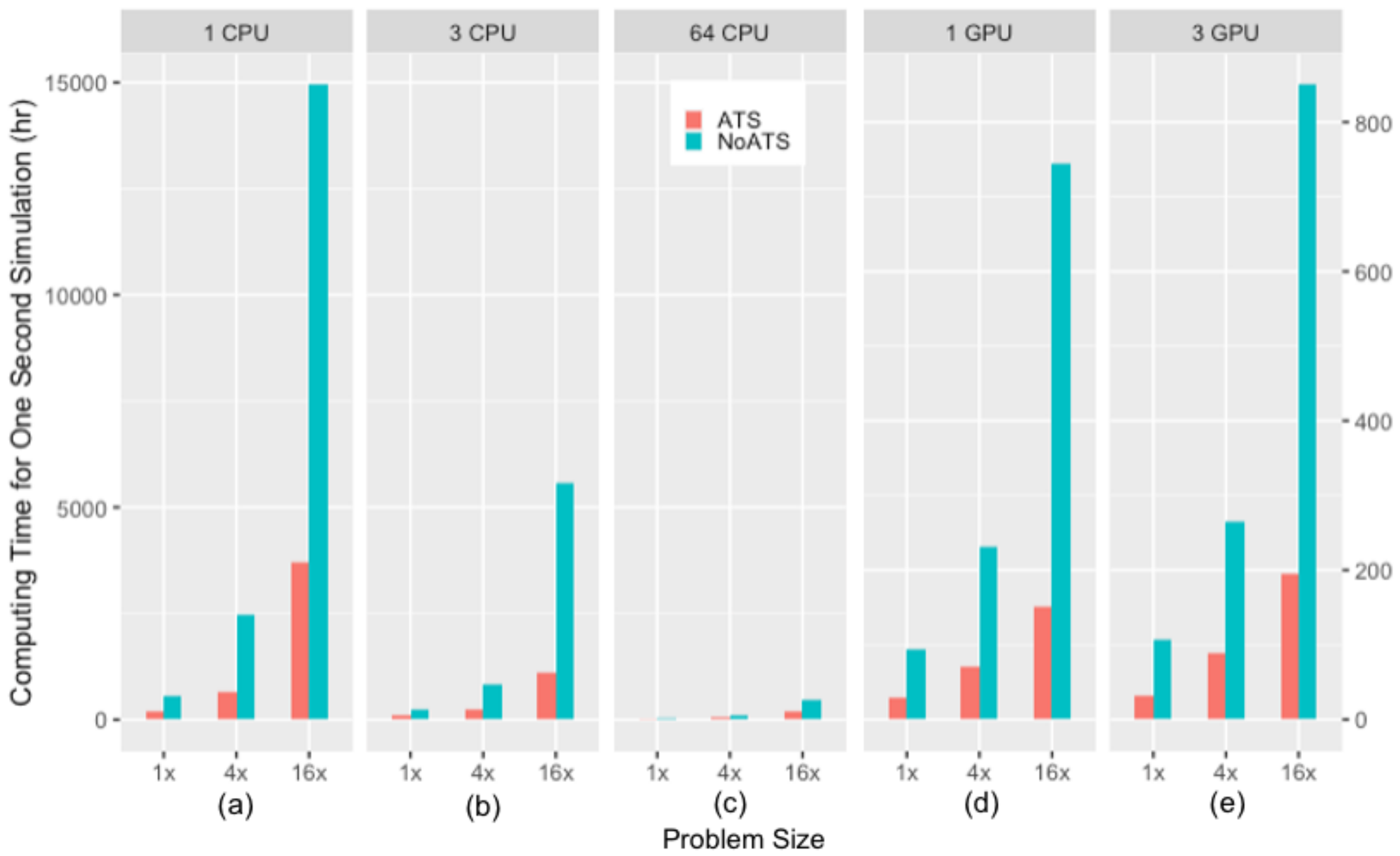}
\caption{Wall clock times for simulations with and without ATS on a single node with CPUs ((a) to (c)) and GPUs ((d) and (e)) with varying problem sizes.}
\label{one_node_ATS-fig}
\end{figure}

\subsubsection{Single node performance}
There are 64 CPU cores and 3 GPUs in a single node of \emph{Blanca} (Table \ref{tab:blancagpu}). Each CPU core is enabled to run 2 threads simultaneously. Wall-clock times on one node (64 and 128 MPI tasks on 64 CPU cores, and 3 MPI tasks on 3 GPUs) were compared and is as shown in Fig.\  \ref{one_node-fig}. The performance of using 3 GPUs is higher than using all CPU cores. The GPU speedup factor with respect to CPU runs is higher at larger problem sizes and is about 4x faster at 32 times the baseline problem size (see Table \ref{tab:problem-sizes}) with 67 million particles. 
We also observe that enabling multiple threads on a single CPU core does not provide appreciable performance improvements.
\subsubsection{Performance improvements with adaptive time-stepping}
We utilize an error controlled adaptive time-stepping method (ATS) \cite{sitaraman2020error} for particle advance where the particle time-steps are dynamically varied based on velocity errors. This allows for increased time-steps when particle dynamics are less stiff thus reducing the total number of particle time-steps per fluid time step. ATS significantly improves the simulation performance for all sizes investigated on both CPU and GPU. The detailed results are plotted in Figure \ref{one_node_ATS-fig} which indicates approximately 4x improvement with using ATS compared to the default constant time step method on both CPUs and GPUs.


\subsection{Parallel performance on multiple nodes}
\begin{figure}
\centering
\includegraphics[width=.95\textwidth]{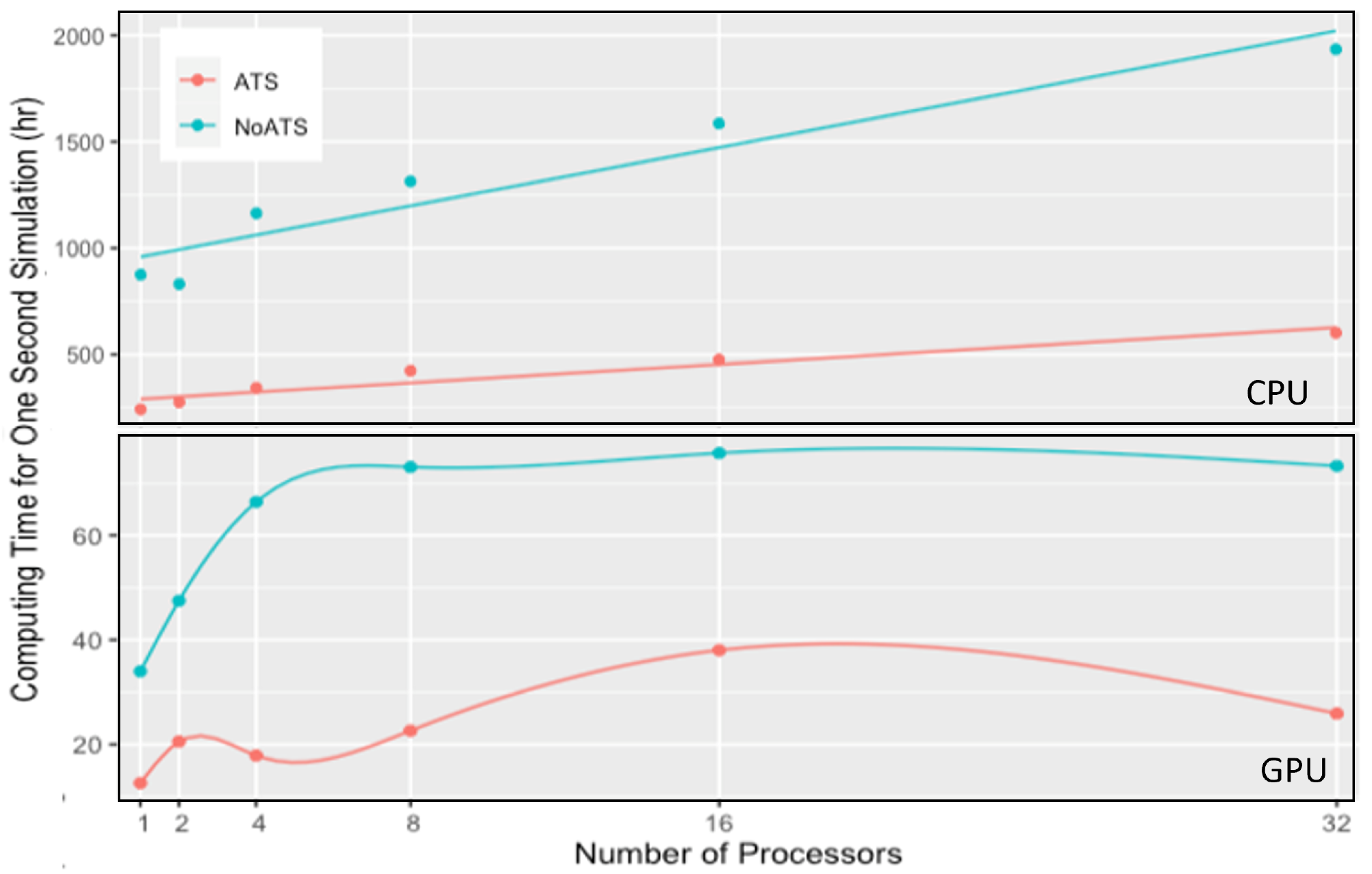}
\caption{Weak Scaling results on CPU and GPU with and without using ATS.  Upper half is the CPU performance while lower half is for GPU cases. Processor here means a single CPU core for the CPU case while it is a single GPU for the GPU case. The number of processors correspond to the problem size factor listed in Table \ref{tab:problem-sizes}. It is also the same as the number of MPI ranks where each MPI task runs on one CPU core or a GPU.}
\label{weak_scaling-fig}
\end{figure}
Figure \ref{weak_scaling-fig} illustrates the weak scaling results on both CPU and GPU with and without ATS. These simulations were performed on \emph{Eagle} supercomputer which has 36 CPU cores and 2 GPUs per node. The wall-clock times for the CPU case tends to increase with increasing number of MPI tasks mainly due to greater memory requirement on a single node and MPI communication costs. On the other hand, the GPU runs tend to stabilize at larger number of MPI tasks (8-32) indicating effective utilization of GPU threads and memory at larger problem sizes. ATS tends to give a consistent performance improvement of approximately 4x across varying number of MPI tasks and problem sizes on both GPUs and CPUs.

\section{Conclusions}
In this work, we presented computational performance comparisons of CFD-DEM simulations performed on 
current CPU and GPU architectures using emerging mulitphase gas-solid solver, MFiX-Exa. A representative fluidized bed case with varying problem sizes (2 to 67 million particles) were simulated and CPU vs GPU performance in serial and parallel paradigms were compared. A single GPU was observed to be about 10 times faster compared to a single CPU core. The use of 3 GPUs on a single compute node was observed to be 4x faster than using all 64 CPU cores. 
We also observed that using an error controlled adaptive time stepping scheme for particle advance provided a consistent 4x speed-up on both CPUs and GPUs. Weak scaling results indicate superior parallel efficiencies when using GPUs compared to CPUs. The GPUs show near perfect parallel efficiencies at large problem sizes while the CPU cases show reduction in efficiencies due to increased memory utilization and communication costs at larger problem sizes.

\section*{Acknowledgments}
This research was supported by the Department of Energy for project titled ``MFIX-DEM enhancement for industry relevant flows" under Grant No. DE-FE0026298. 
The authors acknowledge helpful discussions with their collaborator Peiyuan Liu at University of Colorado, Boulder, Jordan Musser and William D. Fullmer from National Energy Technology Laboratory and Ann Almgren from Lawrence Berkeley National Laboratory 
A portion of this work utilized resources from the University of Colorado Boulder Research Computing Group, which is supported by the National Science Foundation (awards ACI-1532235 and ACI-1532236), the University of Colorado Boulder, and Colorado State University.
A portion of this research was performed using computational resources sponsored by the Department of Energy's Office of Energy Efficiency and Renewable Energy and located at the National Renewable Energy Laboratory.
The U.S. Government retains and the publisher, by accepting the article for publication, acknowledges that the U.S. Government
retains a nonexclusive, paid-up, irrevocable, worldwide license to publish or reproduce the
published form of this work, or allow others to do so, for U.S. Government purposes.




\clearpage
\bibliographystyle{elsarticle-num}
\bibliography{refs}
\end{document}